\documentclass[conference]{IEEEtran}
\IEEEoverridecommandlockouts

\usepackage{cite}
\usepackage{amsmath,amssymb,amsfonts}
\usepackage{algorithmic}
\usepackage[graphicx]{realboxes}
\usepackage{textcomp}
\usepackage[table,xcdraw]{xcolor}
\usepackage{multirow}
\usepackage{array}
\usepackage{tikz}
\usepackage{siunitx}
\usepackage{lscape}
\usepackage{adjustbox}
\usepackage{tabularx}
\usepackage{mleftright}
\usepackage{etoolbox}
\usepackage{xcolor}
\usepackage{amsmath}
\usepackage{amssymb}
\usepackage[normalem]{ulem}
\usepackage{tablefootnote}
\usepackage{verbatim}
\usepackage{graphicx}
\usepackage{booktabs}
\usepackage{capt-of}
\def\BibTeX{{\rm B\kern-.05em{\sc i\kern-.025em b}\kern-.08em
    T\kern-.1667em\lower.7ex\hbox{E}\kern-.125emX}}
\begin{document}

\title{Neurophysiological Analysis in Motor and Sensory Cortices for Improving Motor Imagination
\footnote{{\thanks{This research was supported by the Challengeable Future Defense Technology Research and Development Program through the Agency For Defense Development (ADD) funded by the Defense Acquisition Program Administration (DAPA) in 2024 (No.912911601) was partly supported by the Institute of Information \& Communications Technology Planning \& Evaluation (IITP) grant, funded by the Korea government (MSIT) (No. RS-2019-II190079, Artificial Intelligence Graduate School Program (Korea University)).}
}}
}

\author{
\IEEEauthorblockN{Si-Hyun Kim}
\IEEEauthorblockA{\textit{Dept. of Artificial Intelligence} \\
\textit{Korea University} \\
Seoul, Republic of Korea \\
kim\_sh@korea.ac.kr}

\and

\IEEEauthorblockN{Sung-Jin Kim}
\IEEEauthorblockA{\textit{Dept. of Artificial Intelligence} \\
\textit{Korea University} \\
Seoul, Republic of Korea \\
s\_j\_kim@korea.ac.kr}

\and

\IEEEauthorblockN{Dae-Hyeok Lee}
\IEEEauthorblockA{\textit{Dept. of Brain and Cognitive Engineering} \\
\textit{Korea University} \\ 
Seoul, Republic of Korea \\
lee\_dh@korea.ac.kr}
}

\maketitle

\begin{abstract}
Brain--computer interface (BCI) enables direct communication between the brain and external devices by decoding neural signals, offering potential solutions for individuals with motor impairments. This study explores the neural signatures of motor execution (ME) and motor imagery (MI) tasks using EEG signals, focusing on four conditions categorized as sense--related (hot and cold) and motor--related (pull and push) conditions. We conducted scalp topography analysis to examine activation patterns in the sensorimotor cortex, revealing distinct regional differences: sense--related conditions primarily activated the posterior region of the sensorimotor cortex, while motor--related conditions activated the anterior region of the sensorimotor cortex. These spatial distinctions align with neurophysiological principles, suggesting condition--specific functional subdivisions within the sensorimotor cortex. We further evaluated the performances of three neural network models--EEGNet, ShallowConvNet, and DeepConvNet--demonstrating that ME tasks achieved higher classification accuracies compared to MI tasks. Specifically, in sense--related conditions, the highest accuracy was observed in the cold condition. In motor--related conditions, the pull condition showed the highest performance, with DeepConvNet yielding the highest results. These findings provide insights into optimizing BCI applications by leveraging specific condition--induced neural activations.
\end{abstract}


\section{INTRODUCTION}
Brain--computer interface (BCI) is communicating technology between humans and devices by recognizing their intentions and status \cite{jeong2019classification, he2024brain, lee2020continuous}. BCI technology can be broadly categorized into two types \cite{kim2015abstract}. First, invasive BCI involves the insertion of microelectrodes into the brain to record the action potentials of neurons located at the electrode sites, enabling the extraction of necessary information \cite{edelman2024non}. Owing to this characteristic, invasive BCI offers a high signal--to--noise ratio (SNR), but it has the drawback of requiring surgical operation \cite{lee2022motor}. In contrast, non--invasive BCI refers to techniques that measure the brain’s electrical activity without the need for surgical operation \cite{prabhakar2020framework, wu2024review}. While non--invasive BCI has a lower SNR compared to invasive BCI, it offers the significant practical advantages owing to no need for surgical operation \cite{ahn2022multiscale}. As a result, non--invasive BCI has been utilized for controlling various external devices, including a drone \cite{lee2021design}, a robotic arm \cite{zhou2023shared}, and a wheelchair \cite{rivera2022cca}.

Motor execution (ME) and motor imagery (MI) are types of BCI paradigms. ME paradigm is used to measure brain signals that occur when performing specific movements \cite{lee2020decoding}. In contrast, MI paradigm is utilized to measure brain signals that occur when imagining the movement of muscles used to perform specific movements \cite{ang2016eeg}. MI--based BCI has been instrumental in developing assistive technologies for individuals with motor impairments. Owing to this aspect, ME-- and MI--based BCI systems have utilized electroencephalogram (EEG), which is the most practical method in non--invasive methods \cite{kim2024towards}. Since both ME and MI tasks are motor--related paradigms, EEG signals while conducting ME and MI tasks have the significantly similar features, including spatial features in the sensorimotor cortex \cite{hardwick2018neural}. Ogawa \textit{et al.} \cite{ogawa2022asymmetric} demonstrate the asymmetric representation of the directed functional connectivity while performing ME and MI tasks. They discovered new brain regions related to motor tasks while finding out similarities and differences between ME and MI tasks. Jia \textit{et al}. \cite{jia2023model} proposed the end--to--end deep learning model using the multi--branch spectral--temporal convolutional neural network with the channel attention and the LightGBM for decoding MI-based EEG signals. Their proposed method achieved 0.86 for classifying 2--class MI tasks and 0.74 for classifying 4--class MI tasks. 

The sensorimotor cortex, the main target for motor tasks, is a brain region that processes both sense-- and motor--related information. Recent studies have highlighted the potential for sense--related information to enhance the decoding of motor--related information \cite{flesher2021brain}. Incorporating sensory feedback has been shown to improve the performances of decoding motor--related information, especially in BCI applications. Sense--related information, such as tactile information, can provide valuable real--time information about limb position and movement, which complements motor--related information during motor planning and execution. Moreover, this integration has been found to improve the robustness of BCI systems in dynamic environments. Therefore, combining sense--related information with motor--related information not only enhances decoding performances but also facilitates more intuitive and naturalistic control of external devices.

In this study, we compared and analyzed ME and MI signals while conducting four conditions (i.e., hot, cold, pull, and push) that could engage the sensorimotor cortex. Through this analysis, we explore the differences between sense--related conditions involving the information of the temperature and motor--related conditions involving arm movements. Additionally, by examining the distinctions between ME and MI tasks during these conditions, we demonstrate the potential for a more detailed analysis of the sensorimotor cortex.\\

\begin{figure}[t!]
\centering
\scriptsize
\centerline{\includegraphics[width=\columnwidth]{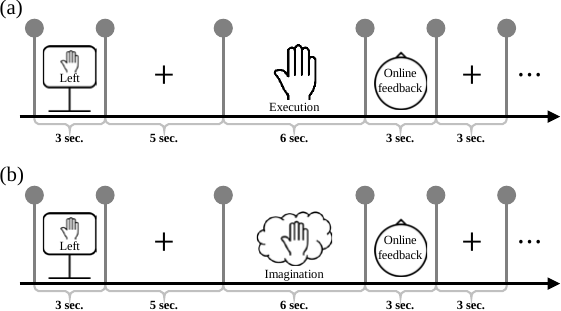}}
\caption{Experimental paradigms for acquiring ME-- and MI--based EEG signals. (a) The paradigm for acquiring ME--based EEG signals. (b) The paradigm for acquiring MI--based EEG signals.}
\end{figure}  

\section{MATERIALS AND METHODS}
\subsection{Subject}
A total of eight healthy subjects (four males and four females, aged 24.9($\pm$1.83)) participated in our experiment. All subjects had no history of neurophysiological disorders. We informed the entire experimental protocol to the subjects, and they consented according to the Declaration of Helsinki. The Institutional Review Board at Korea University (KUIRB--2020--0318--01) reviewed and approved the experimental environment and protocols.\\

\subsection{Experimental Paradigms and Tasks}
We designed the experimental paradigms to acquire ME-- and MI--based EEG signals, as shown in Fig. 1(a) and 1(b), respectively. The experiments of ME and MI were progressed through the same process. We instructed the subjects to perform the ME experiment first and then the MI experiment. Our experimental paradigms had five phases. One of two classes (i.e., left hand or right hand) was displayed over 3 sec. randomly. A fixation cross was provided over 5 sec. to remove the remaining feeling of a cue image and a blank image was represented for 6 sec. The subjects were instructed to perform a task (real movement in the ME experiment and imagination of movement in the MI experiment) when a blank image was displayed. The scalp topography generated by the online feedback system we developed appeared over 3 sec. The online feedback system generates the scalp topography by utilizing EEG signals generated while subjects performed a task in each trial. Through the generated scalp topography, the subjects could confirm whether they performed correctly or not. A 3 sec. video was made using the task performance phase at double speed. After that, a fixation cross appeared for 3 sec. to eliminate the remaining feeling of previous task performance. We acquired 50 trials per class (a total of 100 trials) in each experiment.

\begin{figure}[t!]
\centering
\scriptsize
\centerline{\includegraphics[width=0.4\columnwidth]{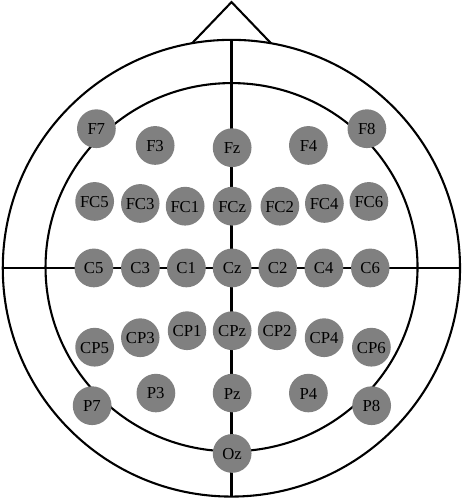}}
\caption{Channel location for acquiring EEG signals.}
\end{figure}  

\begin{figure*}[t!]
\centering
\scriptsize
\includegraphics[width=\textwidth]{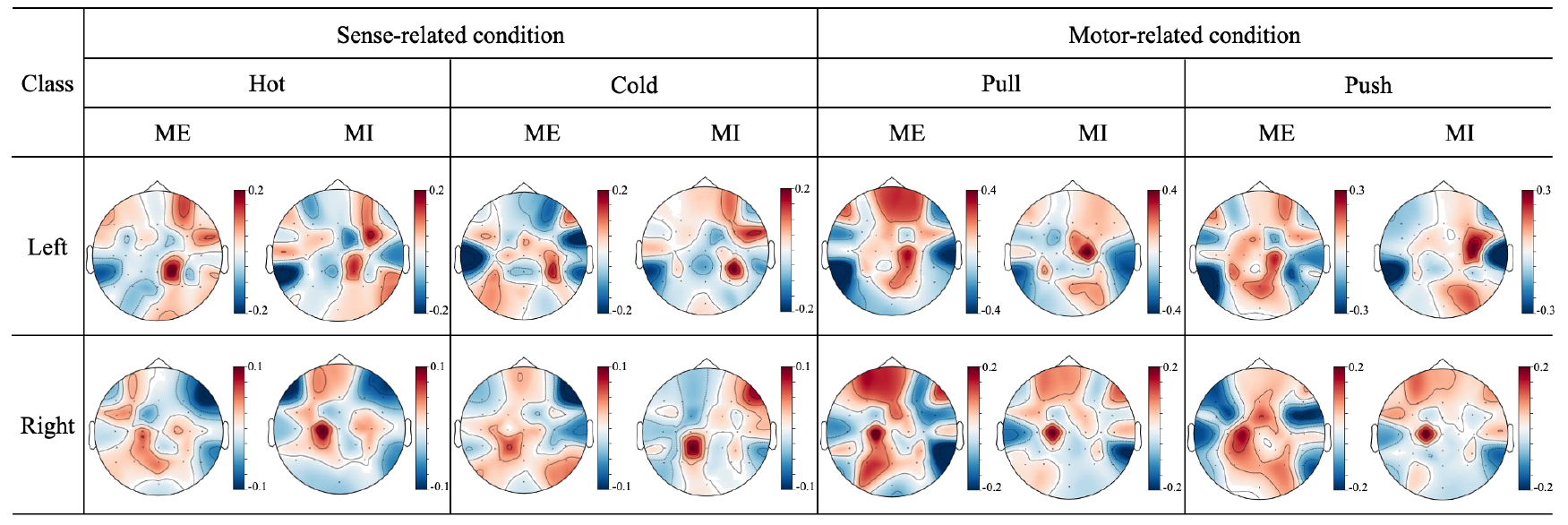}
\caption{Comparison of the sclap topographies on various conditions in ME and MI tasks. We divided conditions into sense--related (hot and cold) and motor--related conditions (pull and push).}
\end{figure*}

In our experiment, subjects were instructed to conduct two tasks (i.e., ME and MI tasks) in each condition (i.e., sense--related and motor--related conditions). We set up the experimental environment including the objects which contain the information of the temperature (i.e., hot and cold) on the desk and resistance bands for conducting motor--related conditions (i.e., pull and push). In the case of sense--related conditions, subjects obtained information on the temperature by grasping the object that corresponds to the class offered in a cue period in the ME experiment. We designed the experimental sequence as follows: \textit{i}) the experiment that offered hot information and \textit{ii}) the experiment that offered cold information. Subjects conducted the MI experiment utilizing obtained information in the ME experiment, without physical movements. In the case of motor--related conditions, we instructed subjects to pull or push the resistance bands by using the hand which corresponds with the class displayed in a cue period for obtaining motor--related information. We instructed the subjects to perform the pull--related experiment first and then the push--related experiment. After finishing the ME experiment, subjects participated the MI experiment by utilizing obtained motor--related information in the ME experiment. Therefore, each subject conducted a total of eight sessions in our experiment.\\

\subsection{Experimental Environment}
In this study, we used the signal amplifier (BrainAmp, Brain Products GmbH, Germany) to measure EEG signals from the subjects. We set the sampling frequency to 250 Hz and applied the 60 Hz notch filter to remove direct current noise. The acquisition of EEG signals involved using 32 EEG channels placed on the subjects' scalps according to the international 10/20 system. The brain region related to ME and MI tasks is the sensorimotor cortex. To focus on extracting meaningful features related to movement, we shifted channels located in regions less related to ME and MI tasks to the sensorimotor cortex, as shown in Fig. 2. Before acquiring EEG signals, we injected conductive gel into the subjects' scalp, ensuring that the impedance of all EEG electrodes was maintained at 10 k$\Omega$ or lower.\\

\subsection{Power Spectral Density (PSD)}
To characterize the power distribution of EEG signals in the frequency domain, we utilized the power spectral density (PSD) to estimate the power. The PSD calculations for each spectral range were performed using the Fast Fourier Transform (FFT). The PSD ($PSD_{f_{2}--f_{1}}$) was computed through the following equation \cite{shin2020prediction}:

\begin{equation}
PSD_{f_{2}--f_{1}} = 10 \cdot \log_{10}{\left(2 \int_{f_{1}}^{f_{2}}|\hat{x}(2\pi f)|^{2} , df\right)},
\end{equation}
where $f_{1}$ and $f_{2}$ represent the lower and upper frequencies, respectively. $\hat{x}(2\pi f)$ was obtained using the FFT. $10 \cdot \log_{10}(\bullet)$ is employed for unit conversion, transforming the values from microvolts to decibels. This approach was used to generate the scalp topography, visualizing the spatial distribution of power across the scalp for each condition.\\

\section{RESULTS AND DISCUSSION}
\subsection{Neurophysiological Analysis from EEG Signals}

Fig. 3 presents the scalp topography analysis of EEG signals for the representative subject, segmented using a 0.5 sec. window across two tasks and four conditions. ME tasks consistently show activation in the sensorimotor cortex, with distinct spatial differences depending on the conditions. Specifically, sense--related conditions exhibit more pronounced activation towards the posterior region of the sensorimotor cortex (sensory cortex), whereas motor--related conditions display activation primarily in the anterior region of the sensorimotor cortex (motor cortex). These patterns align with known neurophysiological distinctions between sensory and motor processing. For MI tasks, the activation patterns also vary based on the types of condition, with sense--related conditions activating regions associated with the sensory cortex, while motor--related conditions engage the motor cortex. This difference in scalp topography suggests that distinct tasks can be used to functionally subdivide the sensorimotor cortex within the EEG domain, highlighting the potential to exploit specific tasks for targeting neurophysiological investigations.\\

\subsection{Performance Evaluation with the Conventional Models} 

We evaluated the performances on three different neural network models--EEGNet \cite{lawhern2018eegnet}, ShallowConvNet \cite{schirrmeister2017deep}, and DeepConvNet \cite{schirrmeister2017deep}--using the average accuracy and standard deviation as metrics, as shown in Table I. These metrics allowed for a comprehensive analysis of the models’ performances across both ME and MI tasks. To ensure the reliability of the results and minimize randomness, we applied the 5--fold cross--validation technique. The results show a comparative analysis of the models’ classification performances in four conditions: hot, cold, pull, and push. For ME tasks, DeepConvNet achieved its highest accuracy in the cold condition, with an accuracy of 0.653$\pm$0.080, outperforming its results in hot and push conditions, where it reached 0.628$\pm$0.081 and 0.594$\pm$0.061, respectively. Across models, the cold condition generally resulted in higher average accuracies compared to the hot condition. In motor--related conditions, EEGNet performed better in the pull condition, achieving 0.596$\pm$0.075, than in the push condition, which had the average accuracy of 0.576$\pm$0.061, indicating variability in condition--specific classification performances. For MI tasks, the performances among the models was more higher than hot and pull conditions in cold and push conditions. ShallowConvNet recorded the highest average accuracy in the pull condition (0.527$\pm$0.794), while EEGNet had higher results in the cold condition. Overall, the average accuracy across all conditions in MI tasks were generally lower than those observed in ME tasks.\\

\begin{table}[t!]
\caption{Comparison of the performances using the conventional models (EEGNet, ShallowConvNet, and DeepConvNet) on various conditions in ME and MI tasks.}
\normalsize
\begin{center}
\renewcommand{\arraystretch}{1.4}{
\resizebox{\columnwidth}{!}{
\begin{tabular}{ccccc}
\cline{3-5}
                                         &                           & EEGNet \cite{lawhern2018eegnet} & ShallowConvNet \cite{schirrmeister2017deep} & DeepConvNet \cite{schirrmeister2017deep} \\ \hline
\multicolumn{1}{c|}{\multirow{4}{*}{ME}} & \multicolumn{1}{c|}{Hot}  & 0.607±0.090 & 0.616±0.089    & 0.628±0.081 \\
\multicolumn{1}{c|}{}                    & \multicolumn{1}{c|}{Cold} & 0.583±0.098 & 0.631±0.089    & 0.653±0.080 \\
\multicolumn{1}{c|}{}                    & \multicolumn{1}{c|}{Pull} & 0.596±0.075 & 0.604±0.741    & 0.624±0.078 \\
\multicolumn{1}{c|}{}                    & \multicolumn{1}{c|}{Push} & 0.576±0.061 & 0.586±0.060    & 0.594±0.061 \\ \hline
\multicolumn{1}{c|}{\multirow{4}{*}{MI}} & \multicolumn{1}{c|}{Hot}  & 0.506±0.084 & 0.513±0.090    & 0.512±0.076 \\
\multicolumn{1}{c|}{}                    & \multicolumn{1}{c|}{Cold} & 0.524±0.093 & 0.530±0.091    & 0.550±0.090 \\
\multicolumn{1}{c|}{}                    & \multicolumn{1}{c|}{Pull} & 0.521±0.075 & 0.527±0.794    & 0.536±0.078 \\
\multicolumn{1}{c|}{}                    & \multicolumn{1}{c|}{Push} & 0.493±0.076 & 0.503±0.080    & 0.507±0.085 \\ \hline
\end{tabular}}}
\end{center}
\end{table}

\section{CONCLUSION}
In this paper, we analyzed EEG signals of ME and MI tasks across four conditions, divided into sense--related (hot and cold) and motor--related (pull and push) conditions. Our results demonstrated that sense--related conditions activated the posterior regions of the sensorimotor cortex, while motor--related conditions engaged the anterior regions of the sensorimotor cortex, supporting the idea of functional subdivisions in the sensorimotor cortex based on condition--specific features. Performance evaluation using EEGNet, ShallowConvNet, and DeepConvNet showed that ME tasks consistently outperformed MI tasks in classification performances. Particularly in the cold condition, where DeepConvNet achieved the highest performance on both ME and MI tasks. Cold and push conditions displayed relatively higher results in sense--related and motor--related conditions, respectively, suggesting that cold and push conditions may provide more discriminative information for decoding. These findings imply that condition--specific activation patterns can be utilized to enhance the performances of BCI systems. Future research will focus on developing advanced algorithms that exploit the distinct activation regions within the sensorimotor cortex to improve the performances and robustness of MI--based system, potentially incorporating adaptive strategies and additional sensory feedback to further refine decoding processes.\\

\bibliographystyle{IEEEtran}
\bibliography{REFERENCE}

\end{document}